# CdTe Spectroscopic-Imager Measurements with Bent Crystals for Broad Band Laue Lenses

N. Auricchio, L. Ferro, J. B. Stephen, E. Caroli, E. Virgilli, O. Limousin, M. Moita, Y. Gutierrez, D. Geoffrey, R. Le Breton, A. Meuris, S. Del Sordo, F. Frontera, P. Rosati, C. Ferrari, R. Lolli, C. Gargano, S. Squerzanti

*Abstract*– In astrophysics, several key questions in the hard X-/soft Gamma-ray range (>100 keV) require sensitivity and angular resolution that are hardly achievable with current technologies. Therefore, a new kind of instrument able to focus hard X and gamma-rays is essential. Broad band Laue lenses seem to be the only solution to fulfil these requirements, significantly improving the sensitivity and angular resolution of the X-/gamma-ray telescopes. This type of high-energy optics will require highly performing focal plane detectors in terms of detection efficiency, spatial resolution, and spectroscopy. This paper presents the results obtained in the project "Technological Readiness Increase for Laue Lenses (TRILL)" framework using a Caliste-HD detector module. This detector is a pixel spectrometer developed at CEA (Commissariat à Energie Atomique, Saclay, France). It is used to acquire spectroscopic images of the focal spot produced by Laue Lens bent crystals under a hard X-ray beam at the LARIX facility (University of Ferrara, Italy).

## I. Introduction

HARD X-/soft gamma-ray astronomy is a crucial window to study the most energetic and extreme events in the Universe. In this spectral window, the previous orbiting telescopes have exploited coded aperture masks with intrinsically high background and limited sensitivity due to the collecting area coincident with the detector area exposed to a source, reducing our view of the Universe. To understand the emission physics of many classes of celestial X-ray sources, three main requirements have to be considered:

- Broad energy band (from fractions of keV to several hundreds of keV);

Manuscript received December 16, 2022.
This work was supported by the Italian Space Agency.
N. Auricchio, J. B. Stephen, E. Caroli and E. Virgilli are with the INAF – OAS Bologna, via Gobetti 93/3, 40129, Bologna, Italy (telephone: +390516398779, e-mail: natalia.auricchio@inaf.it).
L. Ferro and M. Moita are with the Department of Physics and Earth Science, University of Ferrara, via G. Saragat 1, 44122, Ferrara, Italy.
O. Limousin, Y. Gutierrez, D. Geoffrey, R. Le Breton and A. Meuris are with the Department of Astrophysics, CEA Saclay, F-91191 GIF-SUR-YVETTE Cedex, France.
S. Del Sordo and C. Gargano are with the INAF - IASF Palermo, via Ugo La Malfa 153, 90146, Palermo, Italy.
F. Frontera and P. Rosati are with the Department of Physics and Earth Science, University of Ferrara, via G. Saragat 1, 44122, Ferrara, Italy and INAF – OAS Bologna, via Gobetti 93/3, 40129, Bologna, Italy.
C. Ferrari and R. Lolli are with the CNR/IMEM, Parco Area delle Scienze 37/A, 43124 Parma, Italy.
S. Squerzanti is with the National Institute for Nuclear Physics, Ferrara Division, via Saragat 1 - I 44122 Ferrara, Italy.

- High sensitivity, at least two orders of magnitude better than IBIS on board INTEGRAL, at the same energies;
- Imaging capability with sub-arcminutes angular resolution.

In addition, the polarisation measurement capability should be taken into account to study the polarisation status of cosmic sources in the hard X and soft gamma-ray energy range.

The only viable way to fulfil these requirements is to combine different focusing telescopes with complementary passbands.

Thanks to the NASA/NuSTAR mission having on board two co-aligned focusing telescopes operating in the 6–79 keV band, very sensitive ($10^{-8}$ photons cm$^{-2}$ s$^{-1}$ keV$^{-1}$) studies of the hard X-ray sky eventually have been possible.

To extend the energy band from 70 keV up to several hundreds of keV, the Laue lenses, based on diffraction from crystals in transmission configuration, provide a potential technical solution to this problem. The advantage of Laue lenses is that the energy bandwidth can be increased up to 700 keV with a focal length of about 20 m, which is still feasible with a single satellite mission.

Laue lenses can also find interesting and beneficial applications besides the astrophysical field. For example, feasibility studies have been performed in the biomedical engineering field, as Laue lenses can provide a high-resolution image of the radioactivity distribution within a restricted region of the patient's body [1]. Furthermore, Laue lenses can be used for the medical treatment of patients with localised tumours [2].

The present-day work describes the preliminary characterisation of the spectroscopic performance and imaging capability of a Caliste-HD module, realised at CEA (Saclay, France), and reports the PSF images produced by a Laue Lens Germanium bent crystal by using it as a focal plane detector.

## II. Laue Lens principle

In a Laue lens, the photons pass through the full crystal in a transmission configuration, using its entire volume for interacting coherently.

In order to be diffracted, an incoming gamma-ray has to satisfy the Bragg condition, reported in (1), which relates the spacing of the lattice planes, $d_{hkl}$, with the energy of incident photons E and with the angle of incidence $\theta_B$ with respect to the chosen set of planes (hkl):

$$2d_{hkl}\sin\theta_B = \frac{nhc}{E} \quad (1)$$

A Laue lens is made of many crystals in transmission configuration, as mentioned before, disposed in order to concentrate the incident radiation onto a common focal spot. A convenient way to visualise the geometry of a crystal lens is to consider it as a spherical cup covered with crystal tiles that have their diffracting planes perpendicular to the sphere, as shown in Fig 1. The focal spot is on the symmetry axis at a distance f, called the focal length, from the cup, and it is given by (2):

$$f = \frac{R}{2} \quad (2)$$

where R is the radius of the sphere of which the spherical cup is a part.

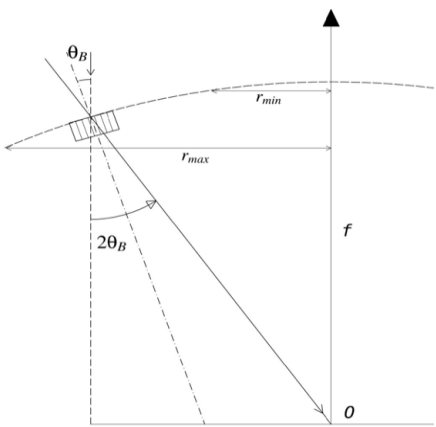

Fig. 1. Laue lens geometry.

Different types of perfect or mosaic crystals can be considered to be implemented in a Laue lens. If the crystal is flat, its diffracted image in the focal plane mainly depends on the crystal size and on the intrinsic spread of the planes of the crystal, which results in a defocusing effect. This defocusing is still present in the bent crystal, but the spot dimension is strongly reduced thanks to the focusing effect dictated by the curvature of the crystal in the bent direction. Instead, no focusing effect is expected in the other direction, as a consequence, the width of the diffracted image has the same size as the crystal. Indeed, in Fig. 2, simulated images are reported when a beam of 20 x 10 mm$^2$ in size and a source of radiation placed at an infinite distance from the target hits a mosaic GaAs (220) crystal tile, which is flat on the left and bent on the right. The focal length is 20 m.

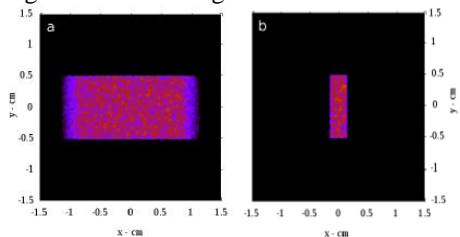

Fig. 2. Diffracted images obtained simulating a beam (20x10 mm$^2$) incident on a mosaic (15 arcsec mosaicity) GaAs (220) crystal tile. The source of radiation is placed at an infinite distance from the target.

For a flat crystal, the diffracted image reproduces the crystal cross-section, while for a bent crystal, it is considerably smaller. In addition, while the diffraction efficiency of mosaic crystals is limited to 50% [3], that of perfect curved crystals can reach 100% [4].

We decided to use bent crystals to extend the energy passband because they accept a wider range of Bragg's angles.

### III. DETECTION SYSTEM

#### A. Caliste HD

Caliste HD [5], realised and provided by CEA (Saclay, Paris), is based on a Schottky CdTe pixelated detector, which is connected to specific Front End electronics, including eight mixed analogue-digital IDeF-X HD ASICs. Each ASIC is equipped with 32 spectroscopic channels and disposed perpendicularly to the detection surface. Fig. 3 shows the Caliste HD module. The CdTe crystal, composed of 256 pixels, is on the top part of the module.

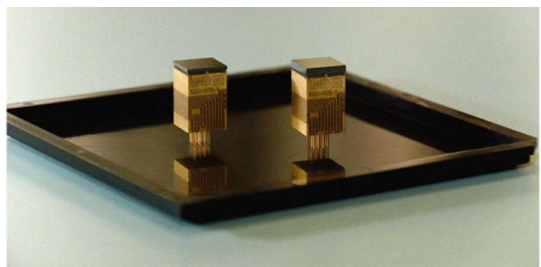

Fig. 3. Caliste HD detector module: the CdTe detector, manufactured by Acrorad (Japan), is bonded on top of it.

The peculiar parameters of the detector are reported in the following table:

TABLE I. CALISTE HD PROPERTIES

| Parameters | Caliste HD |
|---|---|
| Number of Pixels | 16x16 = 256 |
| Pixel pitch | 625 μm (525+100) |
| Guard ring width | 20 μm |
| Thickness | 1 mm |
| Size | 1 cm x 1 cm = 1 cm$^2$ |
| Front End Electronics | IDeF-X HD (32 channels) |
| Number of ASICs | 8 |
| Energy range | 2.5 keV – 1 MeV |
| FWHM at 662 keV | 0.62 % |
| Power consumption | 200 mW |
| Radiation Hardness | ☺ |

Eventually, it is worth mentioning that Caliste HD is already designed for space applications fulfilling radiation hardness, reliability and low-power property requirements.

#### B. Experimental setup

A first detector characterisation was performed at OAS – Bologna laboratory (Fig. 4), where a low-weight portable camera, containing Caliste HD and named Wix Camera, was installed. It is very compact, as illustrated in the pictures of Fig. 5, as it includes the readout electronics, High Voltage module, HV and temperature monitors, power supply converters for the thermoelectric coolers and the Readout board, a Mini Wi-Fi router and an Ethernet connector. The thermoelectric coolers are necessary to cool down the detector in the - 5°C÷ +5°C range.

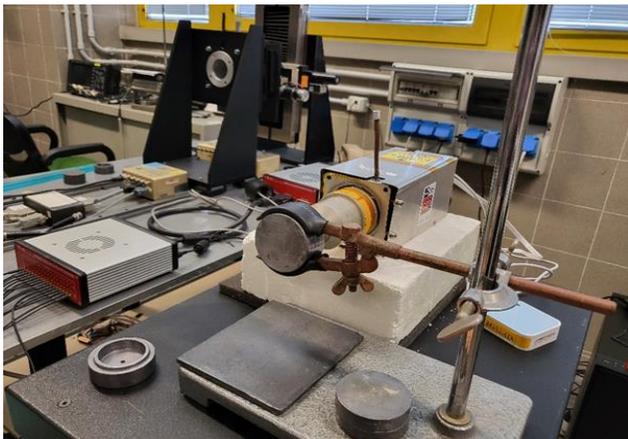

Fig. 4. Experimental setup available at the OAS Bologna premises. The radioactive source, contained in a Pb collimator placed in front of the radiation entrance window, is aligned to the detector with a laser. The detector window is made of 0.5 mm thick Al.

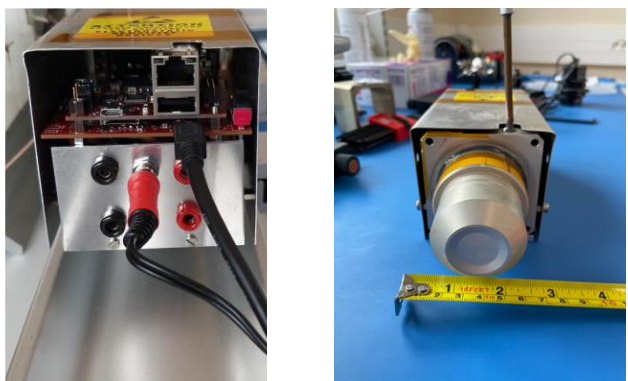

Fig. 5. Portable camera containing the CdTe detector and the related FEE. On the left, the connected Readout and thermoelectric coolers power cables are visible as well the Ethernet and Wi-Fi connectors. The copper tube ensures the hermiticity of the detector head (on the right) filled with Argon.

### C. Quick look and acquisition Software

A short overview of the Quick look SW, also used to acquire at the same time images and calibrated spectra thanks to a calibration table which is loaded each time the application is launched, is shown in the figure below.

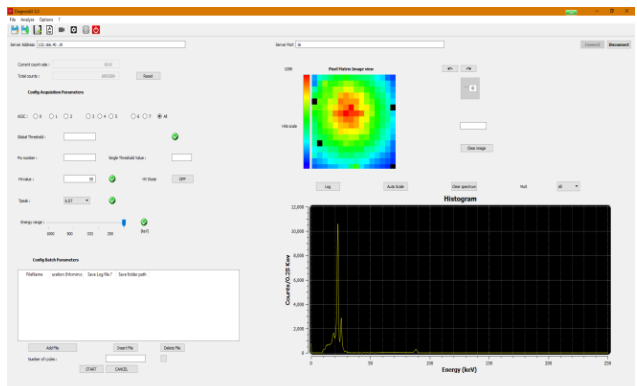

Fig. 6. Quick look and acquisition software supplied with the Wix camera by CEA.

By default, the energy scale of the histogram is already calibrated in keV. Specific configurations can be loaded by selecting the High Voltage, the gain, the peaking time, and the low threshold. The software nevertheless allows passing from the calibrated data to those not calibrated.

### D. Spectroscopic characterisation

By using the experimental setup described in the previous section, we have performed a complete spectroscopic characterisation of Caliste HD by determining the pixel response to irradiation with different calibration sources such as $^{241}$Am, $^{57}$Co, $^{109}$Cd, $^{133}$Ba and $^{137}$Cs to cover the entire dynamic range, and loading all the available configurations listed in Table II.

TABLE II. CAMERA OPERATIONAL CONFIGURATIONS

| Parameter | Value |
|---|---|
| Global Threshold | 20, 30, 50 |
| Gain | 0 (0÷1 MeV) |
| | 3 (0÷250 keV) |
| High Voltage | 190 V |
| | 245 V |
| Peaking Time | 8 (6.07 μs) |
| | 5 (4.06 μs) |
| | 2 (2.05 μs) |
| Sources | $^{241}$Am, $^{109}$Cd, $^{57}$Co, $^{133}$Ba, $^{137}$Cs |

The preliminary results of this spectroscopic characterisation are reported in this section, focusing on the energy resolution obtained by biasing the detector at -245 Volt and applying different threshold values. Fig. 7 compares the spectra acquired by irradiating Caliste HD with $^{241}$Am and setting the peaking time to ~2 and 4 μs.

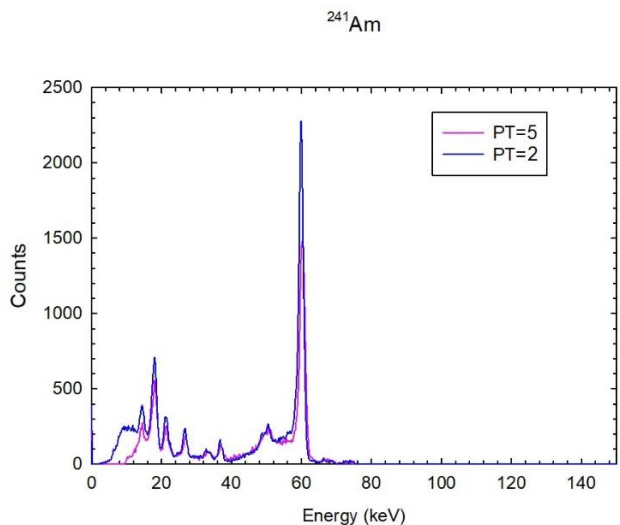

Fig. 7. $^{241}$Am spectrum of all single events acquired with Caliste HD.

Considering all the single events at different Peaking Times, these spectra show an energy resolution of 2.4 % at 60 keV. It is worth pointing out that the lines at low energies are very well separated in this spectrum at 14, 17.9, 21.2, 26.4 keV above all if compared with standard CdTe/CdZnTe detectors.

The other peaks at 36.8 and 33 keV are the CdTe escape peaks due to the escape of the $K_\alpha$ and $K_\beta$ X-rays from tellurium and cadmium. These energies are about 27.5, 26.1, and 23.2 keV. The same behaviour can be observed in the $^{109}$Cd and $^{133}$Ba spectra reported below. In the first spectrum of the $^{109}$Cd source, the peaks at low energy, ~22 and 25 keV, are visible, and in the second one of $^{133}$Ba, the peaks at ~ 31 & 35.4 keV.

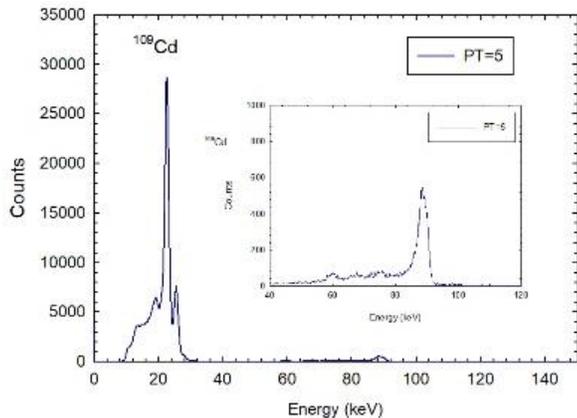

Fig. 8. $^{109}$Cd spectrum of all single events acquired with Caliste HD.

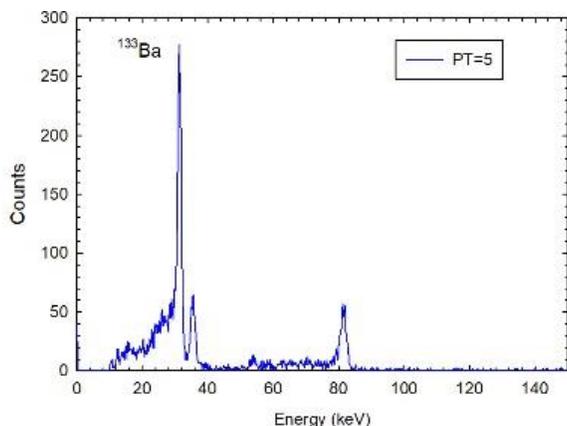

Fig. 9. $^{133}$Ba spectrum of all single events acquired with Caliste HD.

The spectra acquired inside the tunnel at the LARIX facility, which is described in the next section, confirm these results (see Fig. 10). The energy resolution is 2.5% at 60 keV and 2.4% at 81 keV (see Fig. 11).

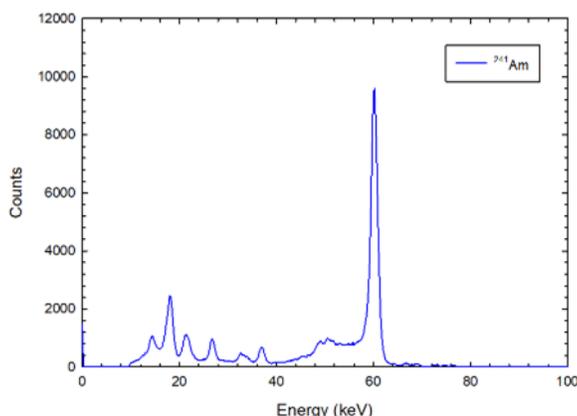

Fig. 10. $^{241}$Am spectrum of all single events acquired with Caliste HD at the Larix facility.

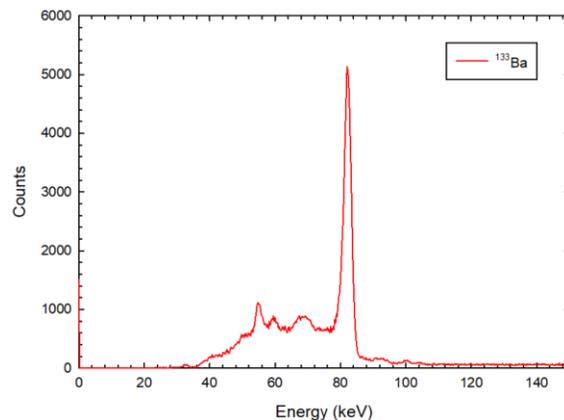

Fig. 11. $^{133}$Ba spectrum of all single events acquired with Caliste HD inside the tunnel at the LARIX facility.

*E. Imaging*

The imaging capabilities have also been evaluated. An example of images acquired at low and high energy is displayed in Fig. 12, irradiating the detector with a collimated source of $^{109}$Cd and an uncollimated source of $^{137}$Cs, from which we can deduce the excellent uniformity of the array.

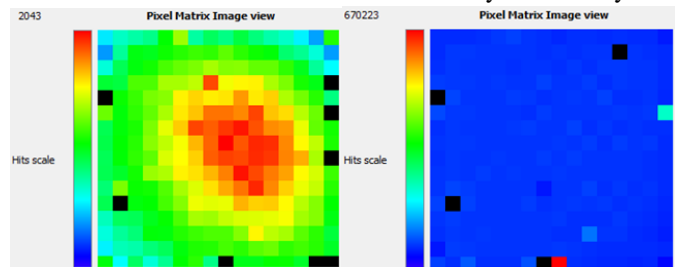

Fig. 12. $^{109}$Cd (left) and $^{137}$Cs (right) images acquired with Caliste HD at the OAS Bologna premises.

IV. LAUE LENS PROTOTYPE MEASUREMENTS

Caliste HD was integrated at the LArge Italian X-ray Facility (LARIX) located in a 100-meter-long tunnel, described in the sketch below, of the Physics and Earth Science Department of the University of Ferrara. The photons coming from an X-ray tube are first collimated, and then they pass through a beamline within which they travel under vacuum (0.1 mbar) to avoid absorption and scattering interactions. The tube is 21 m long and has a diameter of 60 cm. The final beam collimation is performed at the exit window of the beamline in a clean room (class ISO 8) where the crystal assembly apparatus is located. After the diffraction, photons are focused on focal plane detectors.

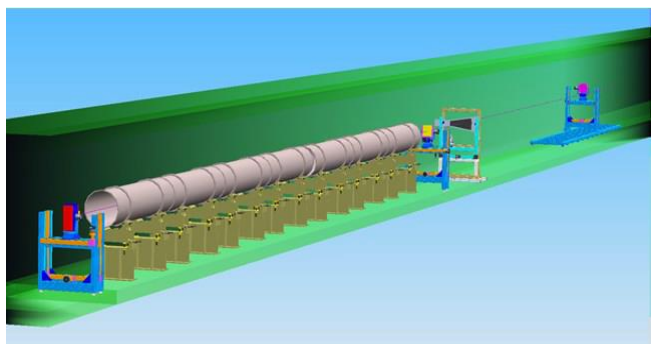

Fig. 13. Drawing of the 100 m long tunnel inside the LARIX facility.

The maximum voltage of the X-ray tube is 320 kV.

Considering the curvature radius of the sample and the divergence of the source of radiation placed, in our experimental setup, at a finite distance inside the facility, the maximum focusing effect is expected at a separation distance between the crystal and the detector of 11.5 m, not a 20 m which is the focal length when the celestial source to be observed in the sky is at an infinite distance.

In the first configuration, the LARIX has been supplied with two focal plane detectors, an imager and a spectrometer, one above the other, held by a pedestal which can be moved along the beam direction (x) and translated in the (y, z) plane. The imager is a Flat Panel X-ray detector based on a CsI(Tl) scintillator, whose fluorescence light is collected from an array of photodiodes on which it is directly deposited. It provides a position resolution of 200 μm. Instead, the spectrometer is based on a cooled High Purity planar Germanium (HPGe) detector with a very good energy resolution (see Fig. 14 on the right). To improve the experimental setup, the spectro-imager Caliste HD (see Fig. 14 on the left) was placed in the focal plane position, located at 11. 5 m from the crystal, and used to test Germanium samples provided by IMEM (Parma). One of these is shown in Fig.15, held in the hexapod and ready to be tested. Table III gives its characteristics.

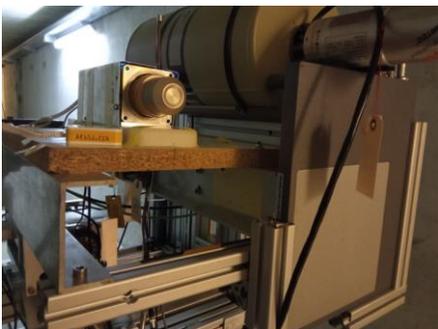

Fig. 14. Caliste HD in the focal plane position located at 11. 5 m from the crystal close to the previous focal plane detectors.

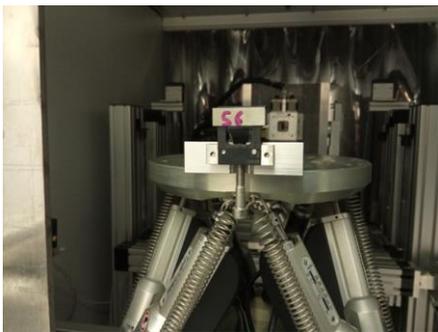

Fig. 15. Ge crystal grabbed by the hexapod available at LARIX to correctly orient each crystal tile to the proper Bragg's angle on the Laue Lens frame under the control of the X-ray pencil beam.

TABLE III. CRYSTAL TILE CHARACTERISTICS

| Material | Germanium |
| --- | --- |
| Diffraction planes | 022 |
| Size | 30 mm x 10mm x 1.6 mm |
| Curvature radius | 39.3 m |

A lapping procedure has been set up by IMEM in order to obtain the desired radius of curvature (~40 m). Indeed, 83 samples with a curvature radius from 37.6 m to 42.3 m were produced, complying with the required accuracy of 2 m, as illustrated in the diagram of Fig. 16.

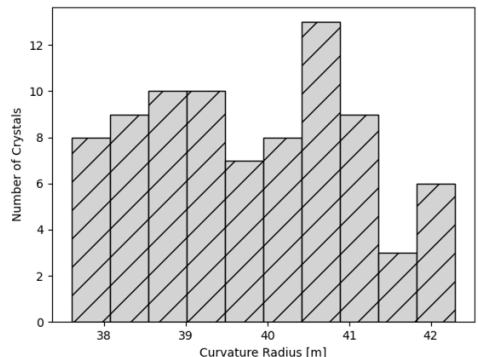

Fig. 16. Distribution of the curvature radius of 83 bent Ge (220) crystals.

### A. First Image and PSF dimensions

To acquire the image of the diffracted X-ray beam at 130 keV through the bent Ge (220) crystal with a 39.3 m curvature radius, the collimator aperture was set to 10 mm x 10 mm in the central position of the sample. The focusing effect along the radial direction is clearly visible in Fig. 17. The focusing factor is about 30.

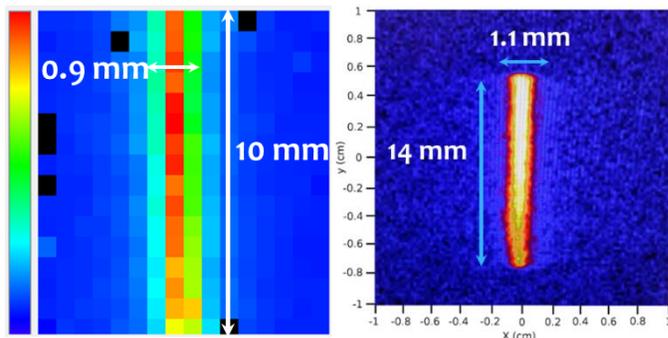

Fig. 17. Measured images with Caliste HD on the left and with the Perkin Elmer Flat Panel, based on a CsI(Tl) scintillator (0.8 mm thick), on the right.

As expected, the divergence of the beam renders the diffracted image from one crystal 14 mm large in the not focusing direction larger than the beam size. At the same time, the Point Spread Function is about 1 mm in the orthogonal direction, thanks to the strong focusing effect. The image on the left is recorded using Caliste HD and compared with one acquired with the Flat Panel imager equipped with a better spatial resolution, shown on the right, obtaining a better result with Caliste HD (0.9 mm).

### B. Broadband

To explore broadband, we have performed two kinds of measurements. Firstly, moving the beam in three positions of the crystal, in steps of 1 cm, to illuminate different parts of it using the same collimator aperture of 10 mm x 10 mm and X-ray beam energy. In the central position, the energy is 132.47±0.06 keV, and the position of the focal spot is different due to the divergence of the beam. The spectrum is reported in

Fig. 18, while the images are in Fig. 19. The focal distance of the crystal is currently at 11.5 m, as already mentioned before, but the lens focal length is at 20 m; thus, the images at 11.5 m will be in focus but shifted between them.

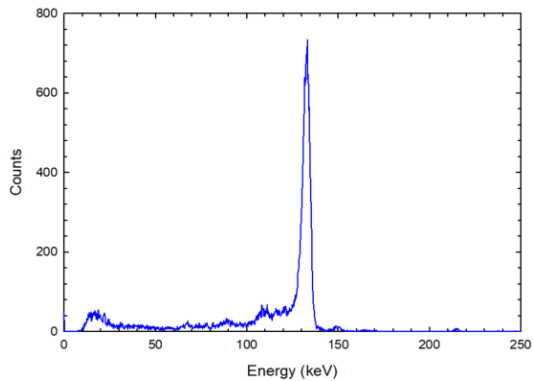

Fig. 18. Spectrum acquired with Caliste HD in the crystal central position at 130 keV. The result of the best fit is 132.47±0.06 keV.

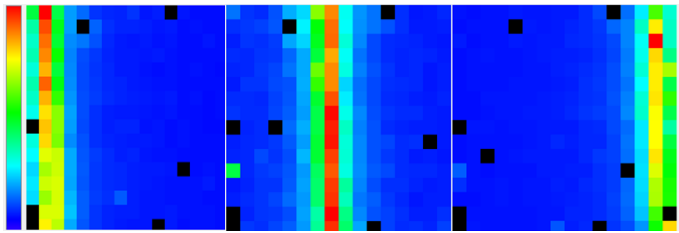

Fig. 19. Diffracted images of the focal spot acquired with Caliste HD in three crystal positions at 130 keV. The images are in focus but slightly shifted between them.

Secondly, carrying out a fine scanning as a function of the position with a step of 2 mm in the central part of the crystal by setting the collimator window to 2 mm x 10 mm to modulate the X-ray beam size incident on the crystal. The acquired image is reported for each position, while only one energy spectrum is shown as an example. The PSF only shifts 1 pixel (0.625 mm)

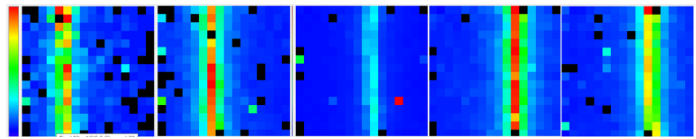

Fig. 20. Focal spot images acquired with the spectro-imager carrying out a fine scanning in the crystal central position in steps of 2 mm at 130 keV.

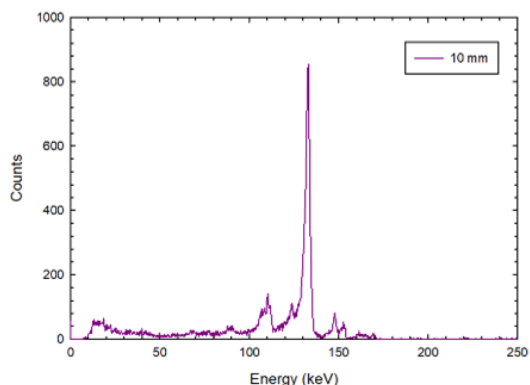

Fig. 21. The related energy spectrum in the last recorded position.

The measured energy band is about 3 keV, as reported in table IV.

TABLE IV. MEASURED ENERGY BAND

| Position (mm) | Energy (keV) |
|---|---|
| 2 | 129.45±0.07 |
| 4 | 128.91±0.10 |
| 6 | 130.60± 0.04 |
| 8 | 131.56± 0.02 |
| 10 | 132.51 ± 0.03 |

The value estimated in the position at 4 mm is probably due to a defect in the crystal.

Scanning of the whole detector with a step of 250 microns and a beam size of 10 mm x 10 mm was performed to reconstruct the position of the beam incident on the sensor. The data analysis is in progress to obtain this position reconstructed by the detector as a function of the real beam position in order to demonstrate the Caliste HD capability to sample the PSF 1 mm wide with a pitch of 0.625 mm. We have already verified that the energy is the same at each position. It is ~132 keV.

The results illustrated in this work are preliminary and will be verified and integrated with other test campaigns at the LARIX facility.

## V. NEXT STEPS

The following steps foresee performing the PSF measurements by using two substrates available at LARIX to investigate the energy passband of the Laue Lens, illuminating each of the three rings of the prototypes. A prototype comprises 11 Ge (220) crystals realised by IMEM (Parma) and bonded on a 5 mm thick quartz substrate with a UV-curable adhesive. The quartz was chosen as petal material, as it is transparent to UV to allow the glue's curing, it has an X-ray transmission coefficient sufficiently high (84% at 150 keV) and a low coefficient of thermal expansion. The substrate is inserted in an INVAR frame, as shown in the photo below (top) [6,7].

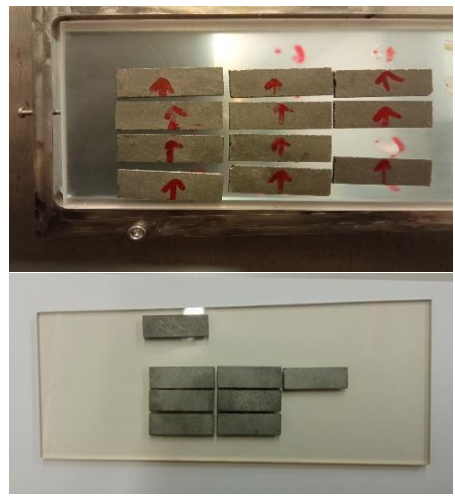

Fig. 22. The Laue lens prototypes are composed of 11 Ge crystals (top) and eight Ge tiles (bottom), each consisting of three rings.

Another prototype was realised by using a 4 mm thick Zerodur substrate, reported in Fig. 22 (bottom). The quartz substrate is bent with the same curvature radius of 40 m as the crystals, while the Zerodur substrate is flat.

The focal spot images recorded using Caliste HD placed in the focal plane of these Laue Lens prototypes will be analysed.

## VI. CONCLUSIONS

An innovative portable camera for X and gamma rays based on a Caliste HD detector was presented. This detector provides very good spectroscopic performances, but the most crucial advantage of this detector for our Laue lens applications is that it is a Spectro imager, able to perform at the same time imaging and energy measurements, to select events based on the energy and position in order to study the structures of the focused images thoroughly and any systematic effects due to the experimental setup.

Currently, Caliste HD is installed at the LARIX facility and used to acquire the PSF of Laue Lens crystals as it is particularly suitable for focal planes of telescopes using focusing optics such as Laue Lens in hard X-rays.

In addition, Caliste HD can be used as an elementary unit for building large detection arrays with limited dead zone thanks to its 4-side buttable geometry.

## ACKNOWLEDGEMENT

This work is partly supported by the AHEAD-2020 Project grant agreement 871158 of the European Union's Horizon 2020 Programme and the ASI-INAF agreement no. 2017-14-H.O "Studies for future scientific missions".